\documentclass[12pt]{article}

\usepackage{amssymb, amsmath, amsthm}
\usepackage{graphicx}
\usepackage{cite}

\newcommand{\td}{\text{d}}

\usepackage{hyperref}

\def\be{\begin{equation}}
\def\ee{\end{equation}}
\def\bea{\begin{eqnarray}}
\def\eea{\end{eqnarray}}

\theoremstyle{definition}

\usepackage[left=2cm,right=2cm,top=2cm,bottom=2cm]{geometry}

\title{\bf Supersymmetric asymptotically locally AdS$_5$ gravitational solitons}
\author{Turkuler Durgut$^a$\footnote{tdurgut@mta.ca } \  and Hari K. Kunduri$^b$\footnote{kundurih@mcmaster.ca } \\ \\
\small \sl $^a$Department of Physics, Mount Allison University \\ \small \sl  Sackville, NB E4L 1E4, Canada  \\
\small \sl $^b$Department of Mathematics and Statistics and Department of Physics and Astronomy \\ \small \sl 
McMaster University, Hamilton, ON  L8S 4M1, Canada }

\date{}

\begin{document}
\maketitle
\begin{abstract} We construct supersymmetric gravitational soliton solutions of five-dimensional gauged supergravity coupled to arbitrarily many vector multiplets. The solutions are complete, globally stationary, $1/4$-BPS and are asymptotically locally AdS$_5$ with conformal boundary $\mathbb{R} \times L(p,1)$.  The construction uses an $SU(2) \times U(1)-$invariant ansatz originally used by Gutowski and Reall to construct supersymmetric asymptotically AdS$_5$ black holes.  A subset of these solutions have previously been obtained as supersymmetric limits of a class of local solutions of $U(1)^3$ gauged supergravity found by Chong- Cvetic-Lu-Pope,  and by Lucietti-Ovchinnikov  in their classification of $SU(2)$-invariant solutions of minimal gauged supergravity. 
\end{abstract}

\vspace{.5cm}
\section{Introduction} 
The Eguchi-Hanson-AdS$_5/\mathbb{Z}_p$ solution constructed by Clarkson and Mann is a globally static, $U(1) \times SU(2)$-invariant, geodesically complete vacuum solution of the Einstein equations with negative cosmological constant $R_{ab} = -4\ell^{-2}g_{ab}$ \cite{Clarkson:2006zk, Clarkson:2005qx}.  The geometry is not asymptotically globally AdS but rather to a freely acting discrete quotient of AdS$_5$.  It has a timelike conformal boundary $\mathcal{I} = \mathbb{R} \times L(p,1)$  where the lens space $L(p,1) \cong S^3 / \mathbb{Z}_p$ with $p \geq 3$.  The solution is therefore best interpreted as an asymptotically \emph{locally} AdS$_5$ gravitational soliton.  The underlying space is characterized by the presence of an $S^2$ `bolt' in the interior region. The solution has negative mass and is conjectured \cite{Dold:2017hwr} to have the least mass amongst the space of metrics with the same conformal boundary, in much the same way that the classic AdS soliton, another vacuum solution of the Einstein equations with negative cosmological constant, is expected to have the least mass amongst all metrics with flat toroidal spatial sections on their conformal boundaries \cite{Galloway:2001uv, Galloway:2002ai}. 

Global AdS$_5$, in addition to being the maximally symmetric vacuum solution (and the unique static spacetime with spherical spatial conformal boundary \cite{Wang}) is also the unique maximally supersymmetric solution of five-dimensional gauged supergravity \cite{Gauntlett:2003fk,Gutowski:2004yv}. In contrast the Eguchi-Hanson-AdS$_5/\mathbb{Z}_p$ soliton does not preserve any (local) supersymmetry. It is natural to expect, however, that it belongs to a larger family of solitons which carry charge and angular momenta, which could possibly contain supersymmetric members.  Indeed,  there are static, charged supersymmetric asymptotically locally AdS solitons with toroidal spatial conformal boundary \cite{Anabalon:2022aig}.  As we discuss below, there are also supersymmetric solutions that are asymptotically locally AdS$_5$ in the sense that the spatial conformal boundary has topology $S^1 \times S^2$ \cite{Cassani:2014zwa}. 

As a byproduct of their recent classification of supersymmetric solutions to five-dimensional minimal gauged supergravity with $SU(2)$ symmetry, Lucietti and Ovchinnikov \cite{Lucietti:2021bbh} constructed a family of $U(1) \times SU(2)$-invariant supersymmetric solitons asymptotic to AdS$_5 / \mathbb{Z}_p$ for $p \geq 3$ (the local solutions had been found earlier in the more general $U(1)^3$ supergravity theory, but a global analysis was not carried out \cite{Cvetic:2005zi}).  The main result of ~\cite{Lucietti:2021bbh} was to establish a uniqueness theorem for $SU(2)$-invariant BPS AdS$_5$ black holes, namely, that the Gutowski-Reall family of solutions~\cite{Gutowski:2004ez} exhaust the moduli space of this class of asymptotically globally AdS$_5$ BPS black holes.  This is a remarkable result, given the lack of uniqueness statements for AdS black holes in any dimension beyond the simple spherically symmetric setting. An analogous result for more general families of toric  ($U(1) \times U(1)$-invariant) BPS AdS$_5$ black holes \cite{Chong:2005hr, Kunduri:2006ek}  remains an open problem, although significant progress has been made in the direction \cite{Lucietti:2022fqj}.  A complete classification of classical BPS AdS$_5$ black hole solutions is motivated by the long standing problem to quantitatively reproduce the Bekenstein-Hawking entropy using the AdS/CFT correspondence (see the review \cite{Zaffaroni:2019dhb} for details on the substantial progress made in recent years). 

Analogous classification results for supersymmetric, horizonless soliton spacetimes remain to be addressed, although in minimal gauged supergravity, the situation for $SU(2)$ symmetry is severely constrained \cite{Lucietti:2021bbh}.  The aim of the present work is to construct new $SU(2)\times U(1)$-invariant, supersymmetric and asymptotically locally AdS$_5$ soliton spacetimes of 
five-dimensional gauged supergravity coupled to an arbitrary number of vector multiplets (see \eqref{action} below).  We will show how these solutions contain the regular solutions found in \cite{Lucietti:2021bbh} (i.e. in a particular truncation of the theory when the scalar fields are constants and the Maxwell fields are set equal).  In addition, we compute the conserved charges of these solutions and show that they have strictly negative mass, despite being supersymmetric and have equal and non-zero angular momenta in two independent planes of rotation.  There is also a BPS relation that appears to depend on $p$.  

Imposing the full set of regularity conditions on our general family of solutions is 
more difficult than in the minimal theory, for which there is only a single (zero-parameter) regular solution for fixed $p$.  However, in the special case of $U(1)^3$ gauged supergravity, we exhibit new explicit solutions which are indeed globally regular (local solutions of this type were constructed in~\cite{Cvetic:2005zi} by taking supersymmetric limits of a family of local metrics, although a global analysis was not performed).  The problem of whether there are asymptotically globally AdS$_5$ solitons within our class of solutions (that is, with $p=1$) remains open but we expect that there is enough freedom in the space of solutions to allow for this possibility. 

Our local construction of these solutions is based upon a mild generalization of an $SU(2) \times U(1)$-invariant ansatz originally employed by Gutowski and Reall in their novel construction of the first asymptotically AdS$_5$ BPS black holes of gauged supergravity coupled to an arbitrary number of vector multiplets \cite{Gutowski:2004yv}.  Supersymmetric solutions of gauged supergravity (in regions where the supersymmetric Killing vector field is timelike) can be constructed in a systematic manner by choosing a four-dimensional  K\"ahler `base space'  and then solving a set of coupled PDEs defined on this base space for various fields that are subsequently used to reconstruct the full spacetime metric, scalars, and Maxwell fields.  In particular, for the class of solutions constructed in the present work, guided by the solutions of the minimal theory~\cite{Lucietti:2021bbh}, we start from a natural $SU(2) \times U(1)-$invariant family of K\"ahler metrics and then, assuming the various geometric fields are themselves invariant under this symmetry, we can proceed in a systematic manner to construct a local family of cohomogeneity-one solutions parameterized by various integration constants.  We then examine the regularity conditions required to extend the local metrics to globally regular, asymptotically locally AdS$_5$ solitons.  Note that these solitons, like the Eguchi-Hanson soliton described above, are characterized by an $S^2$ `bolt'. This is in contrast to the novel BPS solutions numerically constructed in \cite{Cassani:2014zwa}, which have conformal boundary $S^1 \times S^2$ and are of `NUT' type (i.e. the spatial hypersurfaces have $\mathbb{R}^4$ topology). 

The gravitational solitons we construct are 1/4 BPS, as are all known asymptotically AdS$_5$ BPS black holes \cite{Kunduri:2006ek}.  This should be contrasted with the asymptotically globally AdS$_5$ multi charged gravitational solitons we previously constructed \cite{Durgut:2021rma} (special cases of these solutions were previously constructed in the minimal theory~\cite{Cassani:2015upa} and in $U(1)^3$ supergravity \cite{Cvetic:2005zi})  which are 1/2-BPS and satisfy a simpler BPS relation which does not include an angular momentum term (see \cite{Cvetic:2005zi} for a discussion of BPS bounds in $U(1)^3$ gauged supergravity) . Those solutions are also cohomogeneithy-one and $SU(2) \times U(1)$-invariant. However,  in the standard decomposition for solutions admitting Killing spinors described above, the K\"ahler base is merely orthotoric and does not inherit the isometry group of the spacetime (this is related to the fact that the Killing spinor fields are not invariant under the symmetry~\cite{Cassani:2015upa}).  As we demonstrated  \cite{Durgut:2021rma} , this second class of solutions must possess an evanescent ergosurface \cite{Gibbons:2013tqa}, which strongly suggests that they are non-linearly unstable due to the stable trapping of null geodesics \cite{Keir:2018hnv, Eperon:2016cdd}.  For the class of solutions constructed in the present work, however, we find no evidence of such ergosurfaces. 

The remainder of this note is organized as follows. Section 2 provides a concise review of the construction of supersymmetric solutions of $U(1)^N$ gauged supergravity and gives details on the derivation of the local solutions. We then impose regularity conditions to produce smooth soliton spacetimes and investigate the asymptotic behaviour of the solutions and compute the conserved charges.  Finally we investigate some special subsets of solutions (in particular $U(1)^3$ `STU' gauged supergravity and minimal gauged supergravity) for which the regularity conditions can be solved explicitly. Section 3 concludes with a brief discussion. 

\label{sec:intro}
\section{Asymptotically locally AdS$_5$ supersymmetric solitons}
\subsection{Supersymmetric solutions to gauged supergravity} Here we briefly review the local construction of supersymmetric solutions to five-dimensional gauged supergravity coupled to $N$ vector multiplets with scalars taking values in a symmetric space. The analysis was originally performed in \cite{Gutowski:2004yv} (see also \cite{Kunduri:2006ek} which uses the same mostly plus signature as used here). The bosonic sector of the theory is governed by the action
\begin{equation}\label{action}
\begin{aligned}
S = \frac{1}{16 \pi} \int \left(  R \star_5 1 - Q_{IJ} F^I \wedge \star_5 F^J - Q_{IJ} \td X^I \wedge \star_5 \td X^J - \frac{1}{6}  C_{IJK} F^I \wedge F^J \wedge A^K  + 2 g^2 V \star_5 1 \right)
\end{aligned}
\end{equation} where the field content $(g, F^I, X^I)$ consists of the spacetime metric, $N$ Maxwell fields $F^I = \td A^I$ ($I = 1 \ldots N$) and the $A^I$ are locally defined $U(1)$ gauge fields, and $N-1$ real scalar fields which are conveniently parameterized by $N$ real scalar fields $X^I$ satisfying the constraint 
\begin{equation}\label{constraint2}
\frac{1}{6} C_{IJK} X^I X^J X^K = 1.
\end{equation}  The $C_{IJK}$ are constants and as a tensor it is totally symmetric, i.e. $C_{IJK} = C_{(IJK)}$ with $I= 1 \ldots N$ (their indices may be raised and lowered with a flat metric). The constant $g$ appearing \eqref{action} is a constant which we will identify below with the inverse AdS length scale.  A particular quadratic combination of scalar fields $X^I$ that arises naturally is
\begin{equation}
X_I = \frac{1}{6} C_{IJK} X^J X^K.
\end{equation}  In terms of these,  the matrix of couplings $Q_{IJ}$ appearing in the action is given by
\begin{equation}
Q_{IJ} = \frac{9}{2} X_I X_J - \frac{1}{2} C_{IJK} X^K.
\end{equation} 
The $C_{IJK}$ are assumed to satisfy the following symmetric space condition
\begin{equation}\label{symmcond}
C_{IJK} C_{J'(LM} C_{PQ)K'} \delta^{J J'} \delta^{KK'} = \frac{4}{3} \delta_{I(L} C_{MPQ)}.
\end{equation}   We note that this condition is satisfied by the $U(1)^3$ gauged supergravity theory that arises from compactification of Type IIB on $S^5$.  More generally, in reductions of eleven-dimensional supergravity on a Calabi-Yau space $Y$, one obtains a five-dimensional ungauged supergravity theory in which the potential $V$ above is the intersection form, $X^I$ and $X_I$ as defined above correspond to moduli parameterizing the size of two- and four-cycles, and $C_{IJK}$ are the intersection numbers of $Y$.  Gauging this theory leads to the theory \eqref{action}.  We refer the reader to \cite{Cacciatori:2003kv} for further discussion. 

For later use, we note that this can be rewritten explicitly as
\begin{equation}
\begin{aligned}
C_{IJK}( C^{JLM} C^{KPQ} + C^{JLP} C^{KMQ} + C^{JLQ} C^{KMP}) =& \delta_{IL} C_{MPQ} + \delta_{IM} C_{PQL}   \\ 
&+ \delta_{IP}C_{QLM} + \delta_{IQ} C_{LMP}.
\end{aligned}
\end{equation}
This condition ensures that $Q_{IJ}$ has an inverse 
\begin{equation}
Q^{IJ} = 2 X^I X^J - 6 C^{IJK} X_K,
\end{equation} 
where, as mentioned above, we make the identification $C^{IJK}:= C_{IJK}$.  This also allows us to invert for $X^I$ in terms of the $X_J$:
\begin{equation}
X^I = \frac{9}{2} C^{IJK} X_J X_K,
\end{equation} 
which then implies
\begin{equation}\label{constraint}
C^{IJK} X_I X_J X_K = \frac{2}{9}.
\end{equation} 
Finally the scalar potential is given by
\begin{equation}
V = 27 C^{IJK} \bar{X}_I \bar{X}_J X_K,
\end{equation} 
where the $\bar{X}_I$ are a set of constants (in the original work on constructing supersymmetric solutions of this theory, the $V$ is expressed in terms of scalars $V_I$ \cite{Gutowski:2004yv} which are proportional to $\bar{X}_I$).   As shown in \cite{Gutowski:2004yv},  the vacuum AdS$_5$ background with radius $\ell = 1/g$ corresponds to $A^I\equiv 0$ and constant scalars $X^I = \bar{X}^I$, and  
\begin{equation}
\bar{X}^I \equiv \frac{9}{2} C^{IJK} \bar{X}_J \bar{X}_K.
\end{equation} 
The special $U(1)^3$ `STU' supergravity theory which arises from the dimensional reduction of Type IIB on $S^5$ corresponds to $N=3$, $C_{IJK} =1$ if $(IJK)$ is a permutation of $(123)$ and $C_{IJK}=0$ otherwise and $\bar{X}^I =1$, or equivalently $\bar{X}_I = 1/3$.  The symmetric space condition~\eqref{symmcond} holds automatically.  For an explicit embedding of this theory into Type IIB supergravity, see \cite{Cvetic:1999xp}.

Given a Killing spinor, one can show that there is a Killing vector field $V$, which is non-spacelike.  In an open spacetime region where $V^2 = -f^2 < 0$ so that $f > 0$ for some function $f$ one can introduce a local chart in which the metric can be decomposed as
\begin{equation}\label{BPSmet}
\td s^2 = -f^2 (\td t + \omega)^2 + f^{-1} h_{ab} \td x^a \td x^b,
\end{equation} 
where $V = \partial/ \partial t$. Supersymmetry implies that the 4d metric $h$ is K\"ahler with K\"ahler form $J$, and the orientation of the base space $B$ is chosen so that $J$ is anti self dual $\star_4 J = -J$. We choose the 5-form $(\td t + \omega) \wedge \td \text{vol}(h)$ to have positively oriented in the full spacetime.  In the following we summarize the necessary and sufficient conditions on the metric, Maxwell fields, and scalar fields to be a supersymmetric solution of the supergravity field equations \cite{Gutowski:2004yv}. We emphasize that apart from these two requirements, no further conditions have yet been made. In Section 2..2 below we will restrict attention to solutions that are invariant under $SU(2) \times U(1)$ symmetry. 

The Maxwell field can be expressed in the form
\begin{equation}\label{BPSMaxwell}
F^I = \td \left[ X^I f(\td t + \omega) \right] + \Theta^I - 9 g f^{-1} C^{IJK} \bar{X}_J X_K J,
\end{equation} 
where the $\Theta^I$ are self-dual two-forms on $B$, and we must have
\begin{equation}\label{trTheta}
X_I \Theta^I = -\frac{2}{3} G^+,
\end{equation} 
and $G^\pm$ is the (anti-)self dual two-form with $\star_4 G^\pm = \pm G^\pm$, defined as
\begin{equation}
G^\pm = \frac{1}{2} f ( \td \omega \pm \star_4 \td \omega).
\end{equation} 
and $\star_4$ refers to the Hodge dual with respect to $(B, h)$.  This can be inverted so that 
\begin{equation}
\td\omega = f^{-1} (G^+ + G^-).
\end{equation}
Since $(B,h,J)$ is K\"ahler, we can define the Ricci two-form
\begin{equation}
\mathcal{R}_{ab} = \frac{1}{2} R_{ab cd} J^{cd}.
\end{equation} 
Supersymmetry implies that $\mathcal{R} =\td P$ where $P$ is the one-form
\begin{equation}
P = 3 g \bar{X}_I \left( A^I - f X^I \omega\right).
\end{equation} 
This determines completely the function $f$ as 
\begin{equation}\label{BPSf}
f = - \frac{108 g^2}{R} C^{IJK} \bar{X}_I \bar{X}_J X_K,
\end{equation} 
and the following condition holds
\begin{equation}\label{ricciform}
\mathcal{R} - \frac{R}{4} J = 3 g \bar{X}_I \Theta^I.
\end{equation}
All these conditions are necessary and turn out to be sufficient to guarantee and the existence of a supercovariantly constant spinor \cite{Gutowski:2004yv}.  All the field equations are satisfied provided $\td F^I =0$ (which is automatically true if we specify potentials), and the Maxwell equations 
\begin{equation}
\td (Q_{IJ} \star  F^J) = -\frac{1}{4} C_{IJK} F^J \wedge F^K
\end{equation} are satisfied.
The Bianchi identity and the Maxwell equation respectively reduce to the following equations on the base space
\begin{equation}\label{BianchiF}
\td\Theta^I = 9g C^{IJK} \bar{X}_J \td (f^{-1} X_K) \wedge J, 
\end{equation}  
and
\begin{equation}\label{Maxwell}
\begin{aligned}
\td \star_4 \td (f^{-1} X_I) =& -\frac{1}{6}C_{IJK}\Theta^I \wedge \Theta^J + 2g\bar{X}_I f^{-1} G^- \wedge J  \\ 
&+ 6 g^2 f^{-2}(Q_{IJ} C^{JMN}\bar{X}_M \bar{X}_N + \bar{X}_I X^J \bar{X}_J) \mathrm{dvol}(h).
\end{aligned}
\end{equation}  

\subsection{Derivation of the local solution}
In their original construction of $SU(2) \times U(1)$-invariant BPS black holes, Gutowski and Reall considered a natural ansatz in which the K\"ahler base itself is assumed to admit an isometric action of $SU(2) \times U(1)$, which is then naturally inherited by the full spacetime \cite{Gutowski:2004yv}. This is in contrast to the soliton solutions considered in our previous work~\cite{Durgut:2021rma} for which the full spacetime admits the full $SU(2) \times U(1)$ symmetry group, but when decomposed into the form \eqref{BPSmet}, the associated K\"ahler base is merely (ortho)toric.  

We take as K\"ahler metric $(h,J)$ the following $SU(2) \times U(1)$-invariant metric
\begin{equation}\begin{aligned}
h &= \frac{\td r^2}{V(r)} + \frac{r^2}{4} (\td \theta^2 + \sin^2\theta \td \phi^2) + \frac{r^2 V(r)}{4} (\td \psi + \cos \theta \td \phi)^2, \\ J &= \td \left(\frac{r^2}{4} (\td \psi + \cos \theta \phi) \right) \end{aligned}
\end{equation}  where $V(r)$ is an arbitrary smooth function. The full $SU(2) \times U(1)$ case is achieved if we choose $\theta \in (0, \pi), \psi \in (0, 4\pi/p), \phi \in (0, 2\pi)$ where $p \in \mathbb{N}$.  It is natural to write the metric in terms of the right-invariant one-forms 
\begin{equation}
\sigma_1 = \sin \psi \td \theta - \cos \psi \sin \theta \td \phi, \quad \sigma_2 = \cos \psi \td \theta + \sin \psi \sin \theta \td \phi, \quad \sigma_3 = \td \psi + \cos \theta \td \phi
\end{equation} In particular the K\"ahler form is simply
\begin{equation}
J = \td \left[\frac{r^2}{4} \sigma_3 \right].
\end{equation} The right-invariant 1-forms $\sigma_i$  obey 
\begin{equation}
\td \sigma_i = -\frac{1}{2} \epsilon_{ijk} \sigma_j \wedge \sigma_k.
\end{equation} They are invariant under the `left-invariant' Killing fields $R_i$
\begin{equation}\begin{aligned}
R_1 &= - \cos\theta \cos \phi \partial_\phi - \sin \phi \partial_\theta + \frac{\cos \phi}{\sin\theta} \partial_\psi , \\
R_2 &= - \cot\theta \sin \phi \partial_\phi + \cos \phi \partial_\theta + \frac{\sin \phi}{\sin \theta} \partial_\psi, \\
R_3 & = \partial_\phi
\end{aligned} \end{equation} and in addition the particular quadratic combination of the $\sigma_i$ appearing in $h$ is invariant under the $U(1)$ generator $L_3 = \partial_\psi$.  The simplest spacetime solution resulting from this class of K\"ahler bases is AdS$_5$ itself, which has $p=1$, vanishing Maxwell fields $F^I =0$, constant scalars $X^I = \bar{X}^I$, and 
\begin{equation}
f = 1, \qquad \omega = \frac{r^2}{2\ell} \sigma_3, \qquad V = 1 + \frac{r^2}{\ell^2}.
\end{equation} where $\ell = g^{-1}$ is the AdS$_5$ length scale normalized so that $R_{ab} = -4 \ell^{-2} g_{ab}$. For $p > 1$, one obtains an AdS$_5$ orbifold, which is singular at the fixed point of the symmetry group at $r=0$. 

Returning to the general metric,  the scalar curvature is determined by $V(r)$ and its derivatives: 
\begin{equation}\label{scalarcurv}
R_h = -\frac{8 (V-1) + 7 r V' + r^2 V''}{r^2}, 
\end{equation} and the Ricci form can be written
\begin{equation}
 \mathcal{R} = \td \left[ P(r) \sigma_3 \right],
\end{equation} with 
\begin{equation}\label{Riccipot}
P(r) = -\frac{1}{4} (r V' + 4 (V - 1)).
\end{equation}  Choose now an orthonormal frame basis for $(B,h)$
\begin{equation}
E^1 = \frac{\td r}{\sqrt{V}}, \quad E^2 = \frac{r \td \theta}{2}, \quad E^3 = \frac{r \sin \theta \td \phi}{2}, \quad E^4 = \frac{r \sqrt{V}}{2} (\td \psi + \cos \theta \td \phi)
\end{equation} and an associated orthonormal frame for the spacetime: $e^0 = f(\td t + \omega)$, $e^i = f^{-1/2} E^i$.   The K\"ahler form can be expressed as
\begin{equation}
J = E^1 \wedge E^4 - E^2 \wedge E^3
\end{equation} which demonstrates that it is anti self dual, i.e. $\star_4 J = - J$.  In order for the full spacetime to inherit the full $SU(2) \times U(1)$ isometry, it is natural to search for solutions $\omega$ of the form $\omega = \omega_3 (r) \sigma_3$.  We then have
\begin{equation}
\td \omega = \frac{2 \omega_3 ' }{r} E^1 \wedge E^4 - \frac{4 \omega_3}{r^2} E^2 \wedge E^3, \qquad
\star_4 \td \omega = \frac{2 \omega_3 ' }{r} E^2 \wedge E^3 - \frac{4 \omega_3}{r^2} E^1 \wedge E^4. 
\end{equation} This gives
\begin{equation}
G^+ =  r f \left[ r^{-2} \omega_3\right] ' (E^1 \wedge E^4 + E^2 \wedge E^3 ), \qquad G^- = \frac{f}{r^3} \left[ r^2 \omega_3\right] ' (E^1 \wedge E^4 - E^2 \wedge E^3)
\end{equation} which are manifestly self-dual and anti self-dual respectively. The Maxwell fields are also assumed to be invariant under $SU(2) \times U(1)$, which allows us to parameterize it in the form
\begin{equation}
A^I = X^I e^0 + U^I \sigma_3
\end{equation} where the $U^I = U^I(r)$ are smooth functions of $r$ to be determined.  The scalar fields are also assumed to be single-variable functions, namely $X^I = X^I(r)$.  Now using \eqref{BPSf} and \eqref{scalarcurv} yields
\begin{equation}
f = \frac{108 g^2 r^2 C^{IJK} \bar{X}_I \bar{X}_J X_K}{8(V-1) + 7 r V' + r^2 V''} .
\end{equation} The Maxwell fields are then
\begin{equation}
F^I = \td A^i = \td (X^I e^0) + \frac{\td U^I}{\td r} \td r \wedge \sigma_3 - U^I \sigma^2 \wedge \sigma^3
\end{equation} where $\sigma^2 \wedge \sigma^3 = \sin\theta \td \theta \wedge \td \phi$.  Comparing this with the general formula for $F^I$ \eqref{BPSMaxwell} allows us to solve for the self-dual 2 forms $\Theta^I$.  Taking the self dual and anti self dual parts of the resulting equation gives two conditions
\begin{align}
\Theta^I &= r \frac{\td}{\td r} \left( \frac{U^I}{r^2} \right) (E^1 \wedge E^4 + E^2 \wedge E^3), \\
\frac{f}{r^3}  \frac{\td}{\td r} \left( r^2 U^I \right) &= -9 g C^{IJK} \bar{X}_J X_K. \label{UIODE}
\end{align} These two equations are sufficient to guarantee that
\begin{equation}
\td \Theta^I = 9 g C^{IJK} \bar{X}_J \td (f^{-1} X_K) \wedge J
\end{equation} which from \eqref{BPSMaxwell} is equivalent to the Bianchi identity $\td F^I =0$.  Now using \eqref{trTheta} yields
\begin{equation} \label{w3ODE}
f^{-1} X_I \frac{\td}{\td r} \left( \frac{U^I}{r^2}\right) = - \frac{2}{3} \frac{\td}{\td r} \left( \frac{w_3}{r^2} \right)
\end{equation} To make progress we will seek a solution for the scalars $X_I$ of the same form as for the $SU(2)\times U(1)$-invariant BPS black holes \cite{Gutowski:2004yv}:
\begin{equation}\label{finvX}
f^{-1} X_I = \bar{X}_I + \frac{q_I}{r^2}, 
\end{equation} where the $q_I$ are constants. This choice guarantees that in the asymptotic region $r \to \infty$, $X_I \to \bar{X}_I$ assuming that $f \to 1$, as expected for asymptotically locally AdS$_5$ metrics.  Using the constraints \eqref{constraint2}, \eqref{constraint} we find from \eqref{finvX} that
\begin{equation}\label{fBPS}
f = \frac{r^2}{[F(r)]^{1/3}}, \qquad F(r) = r^6 + \alpha_2 r^4 + \alpha_1 r^2 + \alpha_0
\end{equation} with
\begin{equation}
\alpha_0 = \frac{9}{2} C^{IJK}q_i q_J q_K, \qquad \alpha_1 = \frac{27}{2} C^{IJK} \bar{X}_I q_J q_K, \qquad \alpha_2 = \frac{27}{2} C^{IJK} \bar{X}_I \bar{X}_J q_K. 
\end{equation}  Using the assumption \eqref{finvX} we can integrate \eqref{UIODE} to find
\begin{equation}
U^I = -\frac{9}{2} g C^{IJK} \bar{X}_J \left[ \frac{\bar{X}_K r^2}{2} + q_K \right] + \frac{U_0^I}{r^2}
\end{equation} where $U_0^I$ are constants. We then integrate \eqref{w3ODE} to find
\begin{equation}
w_3 = \frac{g \alpha_2}{2} + \frac{g \alpha_1}{4 r^2} + w_0 r^2 - \frac{3 U_0^I \bar{X}_I}{2 r^2} - \frac{U_0^I q_I}{r^4}
\end{equation} where $w_0$ is another integration constant. Comparing this functional form to that of the minimal solution~\cite{Lucietti:2021bbh} suggests that generically  $U_0^I \neq 0$.  The integration constant $w_0$ will be determined by imposing the Maxwell equation below.   We now arrive at an explicit expression for $\Theta^I$: 
\begin{equation}
\Theta^I = \left(\frac{9 gC^{IJK}\bar{X}_J q_K}{r^2} - \frac{4U_0^I}{r^4}\right) (E^1 \wedge E^4 + E^2 \wedge E^4).
\end{equation}  Substituting this into the condition \eqref{ricciform} produces an ODE for $V(r)$: 
\begin{equation}\label{Vcons}
-2r^2 + 2V r^2 - \frac{3}{4} r^3 V' - \frac{r^4 V''}{4} = 2g^2 \alpha_2 r^2 - 12 g \bar{X}_I U_0^I.
\end{equation} In the minimal theory, a general analysis of supersymmetric solutions was carried out in this symmetry class. It was shown that the soliton solutions (i.e. those with a `bolt') must have a $V(r)$ that has a simple zero at some $r_0 > 0$ (the case of having an event horizon or a NUT solution corresponds allowing allowing $r$ to reach zero with $f$ vanishing or non vanishing respectively).  We therefore assume $V(r)$ takes the factorized form
\begin{equation}
V(r) = \frac{(r^2 - r_0^2)(a_0 + a_1 r^2 + r^4 g^2)}{r^4}
\end{equation} for some constants $(r_0,a_0, a_1)$. Inserting this form of $V(r)$ into \eqref{Vcons} imposes
\begin{align}
a_1 & = 1 + g^2 r_0^2 + g^2 \alpha_2 \label{a1cons}  \\
a_0 & = r_0^2 (1 + g^2 r_0^2 + g^2 \alpha_2) - 6 g \bar{X}_I U_0^I = r_0^2 a_1 - 6g \bar{X}_I U_0^I . \label{a0cons}
\end{align} The remaining necessary and sufficient requirement for a supersymmetric solution is the Maxwell equation \eqref{Maxwell}.  The left hand side is easily found to be
\begin{equation}
\td \star_4 \td (f^{-1} X_I)  = \frac{2 q_I V'}{r^3} = q_I \left(\frac{4 g^2}{r^2} + \frac{4}{r^6}(-a_0 + a_1 r_0^2) + \frac{8 a_0 r_0^2}{r^8} \right) .
\end{equation} The right hand side is significantly more complicated, involving various products of the parameters $U_0^I$ and $q_I$.  We will not give it here, but simply note that it can be expressed as a sum of even powers of $r$. 
%\begin{equation}\begin{aligned}
% -\frac{1}{6} C_{IJK} \Theta^J \wedge \Theta^K  & = -\frac{27g^2}{r^4} C_{IJK} C^{JPQ} \bar{X}_P q_Q C^{KMN} \bar{X}_M q_N - \frac{16}{3 r^8} C_{IJK} U_0^J U_0^K \\ &+ \frac{24g}{r^6} C_{IJK} U_0^J C^{KMN} \bar{X}_M q_N \\
% 2g \bar{X}_I f^{-1} G^- \wedge J & = -\bar{X}_I \left( \frac{4 g^2 \alpha_2}{r^2} + 16 g w_0  + \frac{8 g U_0^I q_I}{r^6} \right) \\
% 6 g^2 f^{-2} Q_{IJ} C^{JMN} \bar{X}_M \bar{X}_N & = \frac{4 g^2}{3} f^{-2} Q_{IJ} \bar{X}^J \\
 %& = g^2 \left[2\bar{X}_I + \frac{6}{r^2}(q_I + \frac{\alpha_2}{3}\bar{X}_I) + \frac{6q_I q_K \bar{X}^K}{r^4}   - \frac{6}{ r^2} C_{IJK} \bar{X}^J C^{KMN} q_M \bar{X}_N  \right. \\ & \left. - \frac{3}{ r^4} C_{IJK} \bar{X}^J C^{KMN} q_M q_N \right] \\
% &= g^2 \left[ 2\bar{X}_I + \frac{4 q_I}{r^2} +  \frac{2q_I \alpha_2}{r^4} - \frac{3}{ r^4} C_{IJK} \bar{X}^J C^{KMN} q_M q_N \right]  \\
% 6 g^2 f^{-2} \bar{X}_I X^J \bar{X}_J &= g^2 \bar{X}_I \left[ 6 + \frac{4\alpha_2}{r^2} + \frac{2 \alpha_1}{r^4} \right]
 %\end{aligned}
%\end{equation}  
Satisfying the Maxwell equations then reduces to matching powers of $r$ (from $r^{-8}$ to $r^0$). In particular, the left hand side has no constant term. This gives the condition
\begin{equation}
  0 = \bar{X}_I ( -16 g w_0 + 8 g^2 ) 
  \end{equation} which fixes the integration constant 
  \begin{equation}
  w_0 = \frac{g}{2} = \frac{1}{2\ell}. 
  \end{equation}  The $r^{-2}$ coefficients agrees automatically and the $r^{-4}$ condition is also automatically satisfied, taking into account the identity
  \begin{equation}
  C_{IJK}C^{JLM} C^{KPQ} \bar{X}_L q_M \bar{X}_P q_Q = -\frac{1}{9} C_{IJK}\bar{X}^J C^{KMQ} q_M q_Q + \frac{2 \alpha_1}{27} \bar{X}_I + \frac{2 \alpha_2}{27} q_I
  \end{equation} which follows from the symmetric space condition.  The $r^{-6}$ and $r^{-8}$ coefficients yield the following constraints: 
  \begin{align}
  (3 \bar{X}_J U_0^J) q_I & = 3 C_{IJK} U_0^J C^{KMN} \bar{X}_M q_N - (U_0^J q_J) \bar{X}_I \label{r6cons} ,\\
  a_0 r_0^2 q_I & = -\frac{2}{3} C_{IJK} U_0^J U_0^K. \label{r8cons}
  \end{align}  We will assume $a_0 , r_0 \neq 0$.   Define
 \begin{equation} u_0^3 : = C_{IJK} U_0^I U_0^J U_0^K.
 \end{equation} Then \eqref{r8cons} imposes
 \begin{equation}\label{ucubed}
 u_0^3 = -\frac{3}{2} a_0 r_0^2 q_I U_0^I.
 \end{equation} Finally  \eqref{r6cons} imposes
 \begin{equation}
 3( \bar{X}_L U_0^L) C_{IJK} U_0^J U_0^K = 3I - u_0^3\bar{X}_I
 \end{equation} where
\begin{equation}
 I = C_{IJK} C^{KMN} C_{NPQ} U_0^J \bar{X}_M U_0^P U_0^Q .
 \end{equation} Rewriting the symmetric space condition \eqref{symmcond} as 
 \begin{equation}\begin{aligned}
 C_{MKN} \left[ C^{KIJ} C^{NPQ} + C^{KIP} C^{NJQ} + C^{KIQ} C^{NJP} \right] &= \delta_{MI} C_{JPQ} + \delta_{MJ} C_{PQI} + \delta_{MP}C_{QIJ} \\
 & + \delta_{MQ} C_{IJP},
 \end{aligned}
 \end{equation} and contracting this with $U_0^J \bar{X}_M U_0^P U_0^Q$ yields
 \begin{equation}
3I =  3 C_{MKN}C^{KIJ} C^{NPQ} U_0^P U_0^Q U_0^J \bar{X}_M = u_0^3 \bar{X}_I + 3( \bar{X}_L U_0^L) C_{IJK} U_0^J U_0^K.
 \end{equation} This identity guarantees that the constraint \eqref{r6cons} is satisfied.  This exhausts the conditions imposed the Maxwell equation. 
 
In summary, we have constructed a local supersymmetric solution that is parameterized by the integration constants $U_0^I$ and the parameter $r_0$.  These determine the parameters $q_I$ implicitly through \eqref{r8cons}, which reads
 \begin{equation}
 3\left[ r_0^2 ( 1 + g^2 r_0^2 + g^2 \alpha_2) - 6 g \bar{X}_K U_0^K \right]q_I = -2 C_{IJK} U_0^J U_0^K
 \end{equation} and recall $\alpha_2$ depends linearly on the parameters $q_I$. 
 
\subsection{Global analysis and conserved charges} 
In the $(t,r,\psi,\theta,\phi)$ coordinate system, the local metric is obviously analytic, and so any potential singularities will occur at the zero sets of $f$ and $V$ \cite{Lucietti:2021bbh}. It turns out that in the minimal theory \cite{Lucietti:2022fqj}  the only regular case with $f=0$ corresponds to the BPS black hole solutions of  \cite{Gutowski:2004ez} and we expect a similar conclusion will hold in the general gauged supergravity theory \eqref{action}.  From \eqref{fBPS}, we see $f$ will vanish at $r=0$ provided $\alpha_0 \neq 0$ (if we allow for $f$ to vanish at $r=0$, then one gets a `NUT' type soliton provided $V(0) =1$ and $V(r)$ is a smooth function of $r^2$, see e.g. the numerical solutions constructed in \cite{Cassani:2014zwa}). To obtain a soliton, we need that $r \geq r_0$ with $V(r_0) = 0$ and $f(r_0) > 0$.   The geometries we have constructed are asymptotically locally AdS$_5$ in the sense that they are conformally compact with a timelike conformal boundary $\mathcal{I} \cong \mathbb{R} \times S^3 / \mathbb{Z}_p$.  This can be seen by introducing a new set of coordinates $(T,R,\hat\psi, \theta,\phi)$ defined by
\begin{equation}\label{AGAdS}
T = t, \qquad \hat \psi = \psi - \frac{2}{\ell} t, \qquad R = \sqrt{ r^2 + \frac{\alpha_2}{3}}.
\end{equation}  In the asymptotic region $R \to \infty$, the solutions take the manifestly static (locally) AdS$_5$ form
\begin{equation} \begin{aligned}
\td s^2 &= -\left(1 + \frac{R^2}{\ell^2} + O(R^{-4})\right) \td T^2 + O(R^{-2}) \td t(\td \hat \psi + \cos \theta \td \phi) + \left(1 + \frac{R^2}{\ell^2} + O(R^{-4}) \right)^{-1} \td R^2 \\ & + \frac{R^2}{4}\left(1 + O(R^{-4})\right) \left((\td \hat \psi + \cos\theta \td\phi)^2 + \td \theta^2 + \sin^2 \theta \td\phi^2 \right).
\end{aligned}
\end{equation} The conformal boundary carries the metric
\begin{equation}\label{boundary}
\td s^2_4 = -\td T^2 + \frac{\ell^2}{4} \left((\td \hat \psi + \cos\theta \td\phi)^2 + \td \theta^2 + \sin^2 \theta \td\phi^2 \right)
\end{equation} Spatial sections are $L(p,1)$ with a round metric provided $\theta \in (0,\pi)$ and the $(\hat\psi, \phi)$ plane has the identifications $(\hat\psi, \phi) \sim (\hat \psi + 4\pi/p, \phi)$ and $(\hat\psi, \phi) \sim (\hat \psi + 2\pi, \phi + 2\pi)$.  Global AdS$_5$ with conformal boundary $\mathbb{R} \times S^3$ corresponds to $p=1$.  Regularity of the metric in the interior will impose restrictions on allowed values of $p$. 

Possible singularities will occur at $f =0$ and $V=0$.  The former situation is characteristic of an event horizon (i.e. the non-spacelike supersymmetric Killing vector field $\partial_t$ becomes null) which is necessarily degenerate.  This will occur if the coordinate $r$ ranges to $r=0$.  As our primary interest is in gravitational solitons, which are globally stationary, we will assume $r_0 >0$ and so $V$ has one root and assume the parameters $\alpha_i$ are chosen so that $f > 0$ for all $r \geq r_0$.  Thus the metric is defined on $\mathbb{R} \times (r_0, \infty) \times L(p,1)$.  At $r = r_0$, the Killing field $\partial_\psi$ degenerates. In a neighbourhood of $r=r_0$, constant time slices have topology $\mathbb{R}^2 \times S^2$ with $r= r_0$ playing the role of the `origin of coordinates' in the $(r, \psi)$ place. We may then compactify the interior region by adding an $S^2$ at $r = r_0$, producing a smooth manifold without boundary. The region $r\geq r_0$ is then geodesically complete (i.e. geodesics that reach $r=r_0$ can be extended again to large values of $r$, as in the behaviour of geodesics near the origin in Euclidean space). Note that this is equivalent to requiring that the K\"ahler metric $h$ has a smooth bolt at $r = r_0$ where the Killing vector field $\partial_\psi$ degenerates.  Regularity of the spacetime metric (removal of Dirac-Misner strings) requires that $\omega(\partial_\psi) =0$ or equivalently $\omega_3(r_0) =0$. This imposes the constraint
\begin{equation}\label{DM}
u_0^3 = 3 a_0 r_0^6 \left[-\frac{g\alpha_2}{4} - \frac{g \alpha_1}{8 r_0^2} - \frac{g r_0^2}{4} + \frac{3 \bar{X}_I U_0^I}{4 r_0^2} \right] .
\end{equation} This guarantees that $\omega$ vanishes as $O(\rho^2)$ where $\rho = \sqrt{ r - r_0}$.  Assuming that $\psi$ is identified with period $4\pi / p$, smoothness at the fixed point of $\partial_\psi$ at $r = r_0$ requires
\begin{equation}\label{regularity}
p = \frac{r_0 V'(r_0)}{2} = a_1 + \frac{a_0}{r_0^2} + g^2 r_0^2
\end{equation}  To see this,  note that the spatial geometry near the bolt ($\rho^2 = r- r_0 \to 0$) is 
\begin{equation}\begin{aligned}
h &= \frac{4 \td \rho^2}{V'(r_0)} + \frac{r_0^2 V'(r_0) \rho^2}{4} (\td \psi + \cos\theta \td\phi)^2 + \frac{r_0^2}{4} (\td \theta^2 + \sin^2\theta \td \phi^2) \\
& = \frac{4}{V'(r_0)} \left( \td \rho^2 + \frac{r_0^2 V'(r_0)^2}{16} \rho^2 (\td \psi + \cos\theta \td\phi)^2 \right) + \frac{r_0^2}{4} (\td \theta^2 + \sin^2\theta \td \phi^2)
\end{aligned}
\end{equation} Thus smoothness as $\rho \to 0$ requires that $\tilde \psi : = r_0 V'(r_0)/4 \psi$ be identified with period $2\pi$. However, since $\psi \sim \psi + 4\pi /p$, we arrive at \eqref{regularity}.  We then obtain
\begin{equation}
a_0 = r_0^2 p - a_1 r_0^2 - g^2 r_0^4
\end{equation} and using \eqref{a0cons} gives
\begin{equation}
a_1 = \frac{1}{2} \left[ p - g^2 r_0^2 \right] + \frac{3 g \bar{X}_I U_0^I}{r_0^2}
\end{equation} and using the formula for $a_1$  \eqref{a1cons} we find
\begin{equation}
g^2 \alpha_2 = \frac{1}{2} \left[p - 2 - 3 g^2 r_0^2 + \frac{6 g \bar{X}_I U_0^I}{r_0^2} \right]
\end{equation} which allows us to solve for $a_0$:
\begin{equation}
a_0 = \frac{r_0^2}{2} \left[p - g^2 r_0^2 - \frac{6g \bar{X}_I U_0^I}{r_0^2} \right].
\end{equation} This then determines $q_I$ via \eqref{r8cons} totally in terms of the integration constants $U_0^I$, $p \in \mathbb{N}$ and $r_0$. With the symmetric space condition we find
\begin{equation}
\alpha_1 = \frac{8( \bar{X}_I U_0^I) u_0^3}{a_0^2 r_0^4}.
\end{equation} Subbing this back into \eqref{DM} gives, using \eqref{ucubed}
\begin{equation}\label{gencons}
 -\frac{3}{2} a_0 r_0^2 q_I U_0^I = -3a_0 r_0^4 \left[ \frac{g r_0^4}{4} - \frac{3}{4} \bar{X}_I U_0^I + \frac{g (\bar{X}_I U_0^I) u_0^3}{a_0^2 r_0^4} + \frac{g}{4} r_0^2 \alpha_2 \right].
\end{equation} This produces is a complicated constraint between $(U_0^I, r_0, p)$ which should determine the allowed values of $p$. We therefore expect smooth soliton solutions with one discrete parameter $p$ and $N-1$ continuous parameters.  We have been unable to analyze all solutions to \eqref{gencons} in generality.  We will proceed below by looking at the special case of the $U(1)^3$ supergravity, and in particular, the special case of minimal supergravity which arises when the three gauge fields are set equal.  

Before doing so, however, we may compute the asymptotic conserved charges associated to our general family of solutions. We emphasize that in general we have not addressed existence of an open set in the space of parameters that actually satisfy the regularity conditions.  To compute the mass, it is convenient to use the Ashtekar-Magnon prescription \cite{AMD}, which assigns a conserved charge to a spacetime with a Killing vector field.  To compute the mass, we use the timelike Killing vector field $\partial_T$ that is non-rotating at infinity. The asymptotic fall-off of the Weyl tensor in the coordinate chart \eqref{AGAdS} as $R \to \infty$ is
\begin{equation}\begin{aligned}
C^T_{\; RTR} &=\frac{1}{R^6}\left[ -4r_0^4\left(1 + \frac{r_0^2}{\ell^2} \right) + 2 \ell^2 \alpha_2 \left(1 - \frac{2 r_0^4}{\ell^4} \right) + 3 \alpha_1 + \frac{4\alpha_0}{\ell^2}  + 6 \ell U_0^I \bar{X}_I \left(3 + \frac{4 r_0^2}{\ell^2} \right) + \frac{16 U_0^I q_I}{\ell}\right] \\ & + O(R^{-8})
\end{aligned}
\end{equation} which yields the mass after a suitable rescaling by the conformal boundary defining function and an integration over $L(p,1)$ spatial boundary at infinity (see \cite{Durgut:2021rma} for details of a similar computation)
\begin{equation}
\mathbf{M}_{\text{AD}} = \frac{\pi}{8p}\left[ -4x^2 \ell^2 \left(1 + x \right) + 2  \alpha_2 \left(1 - 2x^2 \right) +\frac{3 \alpha_1}{\ell^2} + \frac{4\alpha_0}{\ell^4}  + \frac{6U_0^I \bar{X}_I}{\ell} \left(3 + 4x \right) + \frac{16 U_0^I q_I}{\ell^3}\right] 
\end{equation} where we have defined the dimensionless parameter $x:=r_0^2 / \ell^2$ which is a rough measure of the radius of the $S^2$ bubble in AdS length units. 
The angular momentum, computed using a Komar integral with respect to $-\partial_\psi$ gives\footnote{the sign in the prefactor follows the convention used in \cite{Gutowski:2004yv}.}
\begin{equation}\label{J}
\begin{aligned}
\mathbf{J} &= \frac{1}{16\pi} \int_{L(p,1)} \star \td \left[g(-\partial_\psi, \sim)\right]   \\
& = \frac{\pi \ell^3}{p} \left[\frac{1}{8 \ell^6}(2 \alpha_0 + \alpha_1 \ell^2) - \frac{x^2\alpha_2}{4 \ell^2} + \frac{U_0^I q_I}{\ell^5} + \frac{3 U_0^I \bar{X}_I}{4 \ell^3} (1 + 2x) - \frac{x^2}{4}(1 + x) \right].
\end{aligned}
\end{equation} The angular momentum associated to the spatial Killing field $\partial_\phi$ vanishes identically. The soliton spaceitme therefore has equal angular momenta with respect to two orthogonal planes of rotation at infinity.  The soliton also carries electric charge defined by 
\begin{equation}
\mathbf{Q}_I := \frac{1}{8\pi} \int_{L(p,1)} Q_{IJ} \star F^J, 
\end{equation} 
where the integral is taken over the conformal boundary as $R \to \infty$ on a spatial hypersurface defined by $t = T = $ constant.  A computation gives
\begin{equation}
\star F^I = -f^{-2} \star_4 \td (X^I f) + e^0 \wedge \left( X^I f \star_4 \td \omega +  \Theta^I + 9f^{-1} g C^{IJK} \bar{X}_J X_K J \right).
\end{equation}  As $R \to \infty$, we find that, pulled back to an $R = $constant, $T = $constant surface, as $R \to \infty$
\begin{equation}\begin{aligned}
\star F^I &= \left(-\frac{9}{8\ell^2} C^{IJK} q_J q_K - \frac{U_0^I}{2\ell} + \frac{3}{4\ell} U_0^J \bar{X}_J \bar{X}^I + \frac{\alpha_1 \bar{X}^I}{8\ell^2} + \frac{ \alpha_2 \bar{X}^I}{4} - \frac{9}{4} C^{IJK} \bar{X}_J q_K\right. \\ &\left. - \frac{9\alpha_2}{2\ell^2} C^{IJK} \bar{X}_J q_K  + O(R^{-2})\right) \sin\theta \td \psi \wedge \td \theta \wedge \td \phi
\end{aligned}
\end{equation} and clearly as $R \to \infty$,  $Q_{IJ} = \tfrac{9}{2} \bar{X}_I\bar{X}_J - \tfrac{1}{2} C_{IJK} \bar{X}^K+ O(R^{-2})$.  This leads to global electric charges
\begin{equation}
\begin{aligned}
\mathbf{Q}_I &= \frac{2\pi}{p} \left[\frac{3}{2}\left( -\frac{\alpha_1}{8 \ell^2} - \frac{3 U_0^K \bar{X}_K}{4\ell} - \frac{3 \alpha_2  \bar{X}^J q_J}{\ell^2} + \frac{ \alpha_2^2}{2 \ell^2}\right) \bar{X}_I  + \frac{3}{8}\left(1 + \frac{2 \alpha_2}{\ell^2} \right) q_I  + \frac{C_{IJK}\bar{X}^J U_0^K}{4\ell} \right. \\ &+ \left. \frac{9}{16 \ell^2} C_{IJK} \bar{X}^J C^{KMN}q_M q_N \right].
\end{aligned}
\end{equation} A considerable simplification arises if one considers the `total charge'  
\begin{equation}
\bar{X}^I \mathbf{Q}_I = \frac{\pi}{p} \left[\frac{\alpha_1}{ 8 \ell^2} + \alpha_2 \left(\frac{1}{4}  - \frac{\alpha_2}{\ell^2} \right) + \frac{3 U_0^I \bar{X}_I}{4\ell} \right].
\end{equation}

We now turn to studying the existence of solutions which satisfy all the regularity conditions. The above regularity condition \eqref{gencons} is difficult to analyze in the general multicharge theory.  We will focus on the standard $U(1)^3$ supergravity theory obtained by dimensional reduction of a truncation of Type IIB supergravity on $S^5$.  We take $I = i = 1,2,3$ and $\bar{X}_i = 1/3, \bar{X}^i = 1$.   The constants $C_{ijk} = |\epsilon_{ijk}| = 1$ if $(i,j,k)$ are a permutation of $1,2,3$ and zero otherwise.  We set $\mathfrak{q}_i:=3q_i$.  Using these relations one finds $\alpha_0 = \mathfrak{q}_1 \mathfrak{q}_2 \mathfrak{q}_3,  \alpha_1 = \mathfrak{q}_1 \mathfrak{q}_2 + \mathfrak{q}_2 \mathfrak{q}_3 + \mathfrak{q}_1 \mathfrak{q}_3, \alpha_2 = \mathfrak{q}_1 + \mathfrak{q}_2 + \mathfrak{q}_3$ and hence
\begin{equation}
f^{-3} = \left(1 + \frac{\mathfrak{q}_1}{r^2} \right)\left(1 + \frac{\mathfrak{q}_2}{r^2} \right)\left(1 + \frac{\mathfrak{q}_3}{r^2} \right).
\end{equation}  We have from \eqref{r8cons} that
\begin{equation}
\mathfrak{q}_1 = -\frac{4 U_0^2 U_0^3}{ a_0 r_0^2}
\end{equation} and similar expressions for $\mathfrak{q}_2, \mathfrak{q}_3$.   The remainder of the solution is then explicitly given by
\begin{align}
\omega_3 & = \frac{g}{2} \sum_i \mathfrak{q}_i + \frac{g}{4r^2} \left(\mathfrak{q}_1 \mathfrak{q}_2 + \mathfrak{q}_1 \mathfrak{q}_3 + \mathfrak{q}_2 \mathfrak{q}_3 \right) + \frac{g r^2}{2} 
- \frac{1}{2 r^2} \sum_i U_0^i  + \frac{4 U_0^1 U_0^2 U_0^3}{a_0 r_0^2 r^4} \\
A^1 & = \left( 1 + \frac{\mathfrak{q}_1}{r^2} \right)^{-1} (\td t + w_3 (\td \psi + \cos\theta \td \phi)) - \left( \frac{g}{2} \left( r^2 +  \mathfrak{q}_2 + \mathfrak{q}_3 \right) + \frac{U_0^1}{r^2} \right) (\td \psi + \cos\theta \td \phi) \\
X^i & = \left[ f \left(1 + \frac{\mathfrak{q}_i}{r^2} \right) \right]^{-1}
\end{align} with similar expressions for $A^2, A^3$ with the obvious permutations of the charge parameters $\mathfrak{q}_i$.

We now examine the restrictions on the parameters in detail. It is convenient to introduce dimensionless parameters $y_0^i := U_0^i/\ell^3 = U_0^i g^3$ and $x:=r_0^2/\ell^2$.  Firstly, \eqref{a0cons} and \eqref{a1cons} give respectively
\begin{equation}\begin{aligned} \label{ycons}
\sum y_0^i &= \frac{g^2}{2} (r_0^2 a_1 - a_0) ,\\
y_0^1 y_0^2 + y_0^2 y_0^3 + y_0^1 y_0^3 &= \frac{g^4 a_0 r_0^2}{4}(1 + x - a_1) .
\end{aligned}
\end{equation} Squaring the first and subtracting twice the second gives
\begin{equation}\label{ysq}
\sum_{i} (y_0^i)^2 = \frac{g^4}{4} (a_0^2 + a_1^2 r_0^4 - 2 a_0 r_0^2 (1+x)).
\end{equation}  %Note that if we set the charge parameters $y_0^i$ equal, one will recover the constraint on $(a_0, a_1)$ we had in the minimal theory, so that is consistent. 
Now the condition that $\omega_3(r_0) =0$ becomes 
\begin{equation}\label{consSTU}
1= \frac{a_1}{2} + \frac{a_0}{2r_0^2}  + \frac{8 y_0^1 y_0^2 y_0^3}{g^2 a_0 x^3} + \frac{1}{4 x}(1 + x - a_1)^2 -\frac{4}{g^4 a_0^2 x^3} ((y_0^1 y_0^2)^2 + (y_0^2 y_0^3)^2 + (y_0^1 y_0^3)^2 ).
\end{equation} The last term above can be written as
\begin{equation}\begin{aligned}
-\frac{4}{g^4 a_0^2 x^3} ((y_0^1 y_0^2)^2 + (y_0^2 y_0^3)^2 + (y_0^1 y_0^3)^2 ) &= \frac{2}{g^4 a_0^2 x^3}((y_0^1)^4  + (y_0^2)^4 + (y_0^3)^4) \\ & - \frac{g^4}{8 a_0^2 x^3} \left[a_0^2 + a_1^2 r_0^4 - 2a_0 r_0^2(1+x) \right]^2. \end{aligned}
\end{equation} We also have the algebraic identity
\begin{equation}
3y_0^1 y_0^2 y_0^3 = (y_0^1)^3  + (y_0^2)^3 + (y_0^3)^3 + \sum_i y_0^i \cdot \left(y_0^1 y_0^2 + y_0^2 y_0^3 + y_0^1 y_0^3  - \sum_j (y_0^j)^2 \right).
\end{equation} One can replace the right hand sides of these identities with expressions for the charge parameters $y_0^i$ in terms of $(a_0, a_1, r_0)$ using \eqref{ycons} and \eqref{ysq}.   Then, we eliminate $a_1$ in terms of $(r_0, a_0)$ using the regularity condition \eqref{regularity}. Next, one can eliminate any power of $a_0$ higher than 1 by solving for $a_0^2$ using \eqref{ysq}.  Finally one can get a linear equation for $a_0$ by substituting this back into \eqref{consSTU}.  This allows us to determine $(a_0, a_1)$ in terms of the $y_0^i$ and $x$. Carrying this out produces a complicated polynomial equation in $(x, y_0^i)$ (8th-order in $x$) which we will not display here. There does not seem to be a simple way to determine all allowed values of $(p,x)$ from this in general unless additional simplifications are assumed.  We will be able to show that the procedure described above can be carried out in two specific cases.

\subsubsection{Equal charges}
In the case of equal charges $\mathfrak{q}_i = \mathfrak{q}$,  $U(1)^3$ theory will simplify, upon suitable field redefinitions, to the minimal theory and our solutions should reduce to those found in~\cite{Lucietti:2021bbh}.  Then we have immediately from \eqref{a1cons} that
\begin{equation} \label{mina1}
a_1 = 1 + x + 3 g^2 \mathfrak{q}.
\end{equation} Since all the gauge fields $A^i = A$, we should have equal integration constants $U_0^i$. Fix the dimensionless parameter $\mathcal{U}_0  = U_0^I / \ell^3$. Then \begin{equation}
\mathfrak{q} = -\frac{4 \mathcal{U}_0^2}{a_0 r_0^2 g^6}.
\end{equation} We then have from \eqref{mina1} 
\begin{equation}\label{Usq}
\mathcal{U}_0^2 = \frac{1}{12} (1 + x - a_1) g^4 a_0 r_0^2.
\end{equation} On the other hand \eqref{a0cons} gives
\begin{equation}
\mathcal{U}_0 = \frac{g^2}{6} (r_0^2 a_1 - a_0)
\end{equation} and combining these two expressions for $\mathcal{U}_0^2$ gives
\begin{equation}\label{mincons}
a_0^2 - 3a_0 r_0^2 + a_0 a_1 r_0^2 + a_1^2 r_0^4 - 3a_0 r_0^2 x =0.
\end{equation} Note that this is precisely \eqref{ysq} when one sets all dimensionless charge parameters equal, i.e. $y_0^i \equiv \mathcal{U}_0$.  The regularity conditions \eqref{regularity} inserted into \eqref{DM} gives the condition
\begin{equation}
-\frac{6 \mathcal{U}_0^2}{x} + \frac{12 \mathcal{U}_0^4}{x^3 g^2 a_0} + \frac{g^2 a_0 x}{2} - \frac{3 \mathcal{U}_0 a_0}{2 r_0^2} + \frac{4 \mathcal{U}_0^3}{x^3} =0.
\end{equation} This gives an equation quadratic in $a_0, a_1$ (keeping in mind to eliminate the $\mathcal{U}_0^4, \mathcal{U}_0^3$ terms using \eqref{Usq}). Then use the regularity condition to get rid of $a_1$ terms, the constraint \eqref{mincons} to eliminate the $a_0^2$ terms , and one is left with the condition 
\begin{equation}
a_0 = \frac{ 2p^2 - 4 p + 3 + (p-8)x}{g^2(p+1)} .
\end{equation}  Subbing back into \eqref{mincons} produces
\begin{equation}\label{minquad}
27 x^2 - (p-2)(p^2 + 14 p - 5)x + (p-2)^3 p =0.
\end{equation} This equation is identical to the regularity condition derived  in \cite{Lucietti:2021bbh}. For fixed $p$ this yields a quadratic for $x$. The analysis of \cite{Lucietti:2021bbh} demonstrates that the only possible smooth solutions have $p \geq 3$ with the corresponding larger (positive) root $x$ of \eqref{minquad}.  For this class of solutions of the minimal theory, there are no continuous parameters once the topology is fixed by the choice of $p$. Explicitly, the spacetme metric is \eqref{BPSmet} is determined by 
\begin{align}
f &= \left(1  - \frac{r_0^2 + 1 -a_1}{3 r^2}\right)^{-1}  \\
\omega_3 & = \frac{r^2-r_0^2}{2\ell} + \frac{(a_1 - 1)\ell}{2} + \frac{r_0^4 + r_0^2 \ell^2(2 - 5a_1) + 3a_0 \ell^2 + \ell^4(1 - a_1)^2}{12 \ell r^2} \\
& - \frac{(a_0 + a_1 r_0^2)(r_0^2 + (1 - a_1)\ell^2)\ell}{18 r^2} - \frac{\ell(a_0^2 \ell^2 + a_0(a_1 - 3) \ell^2 r_0^2 - 3a_0 r_0^4 + a_1 \ell^2 r_0^4)}{18 r^6},
\end{align} along with the relations
\begin{gather}\label{minimal}
a_1 = p - x - \frac{a_0}{r_0^2}, \qquad \frac{a_0}{r_0^2} = \frac{2 p^2 - 4p + 3 + (p-8)x}{x(p+1)} ,\\
x = \frac{p-2}{54} \left((p^2 + 14p - 5) + (1+ p)\sqrt{(1+p)(25+p)}\right),
\end{gather} and the Maxwell field by
\begin{equation}
F = \td \left[ \frac{\sqrt{3}}{2}  \left[ f ( \td t + \omega) \right] + \frac{\ell}{2\sqrt{3}} P \right]\end{equation} where $P$ is given by \eqref{Riccipot}.  

For these globally smooth and stationary gravitational solitons with conformal boundary $\mathbb{R} \times L(p,1)$, we may compute their conserved charges. The asymptotic integrals are computed in the limit $R \to \infty$ in the asymptotically static $(T, R, \bar\psi, \theta, \phi)$ coordinate chart \eqref{AGAdS}.  For the mass, it is natural to use the Ashtekar-Magnon mass, which yields the result
\begin{equation}\label{ADmass}
\mathbf{M}_{\text{AD}} = -\frac{(p-2)^2 (2p + 5) \ell^2 \pi}{108 p}
\end{equation} where we have used the asymptotically static Killing field $\partial_T$ to define the mass.  Note that since $p \geq 3$, $M_{\text{AD}}$ is strictly negative. This is not unexpected, as these solutions are not asymptotically globally AdS$_5$ and a positive energy theorem does not apply. Indeed, the mass of the vacuum Eguchi-Hanson-AdS$_5/\mathbb{Z}_p$ soliton family of solutions (whose members are also asymptotically locally AdS$_5 / \mathbb{Z}_p$ for $p \geq 3$) is strictly negative.  The angular momentum computed using the Komar integral \eqref{J} is
\begin{equation}
\mathbf{J}  = -\frac{\ell^3  (p-2)^3 \pi}{108 p}.
\end{equation} Finally, the electric charge can be computed from the integral 
\begin{equation}
\mathbf{Q} = \frac{1}{4\pi} \int_{L(p,1)} \star F  = - \frac{\ell^2 (p-2)\pi}{6p\sqrt{3}}
\end{equation} where we have used the formula
\begin{equation} \begin{aligned}
\star F & = \frac{\sqrt{3}}{2}  \left[ \frac{f ' V r^3}{8f^2 } \sin \theta \td \theta \wedge \td \phi \wedge \td \psi + \frac{r f^2 \omega_3 ' }{2} (\td t + \omega_3 \sigma_3) \sin \theta \wedge \td \theta \wedge \td \phi - \frac{2 f^2 \omega_3}{r} \td t \wedge \td r \wedge \sigma_3 \right]  \\
& + \frac{\ell}{2\sqrt{3}} \left[ \frac{ r f P'}{2} (\td t + \omega_3 \sigma_3) \wedge \sin \theta \td \theta \wedge \td \phi - \frac{2 P f}{r} (\td t + \omega_3 \sigma_3) \wedge \td r \wedge \sigma_3 \right] .
\end{aligned}
\end{equation} Using these definitions, we arrive at the BPS-type relation ($p \geq 3$)
\begin{equation}
\mathbf{M}_{\text{AD}} = \frac{\sqrt{3}(p-2) \mathbf{Q}}{2}+\frac{2 \mathbf{J}}{\ell}.
\end{equation} One can of course define the mass for a family of non-supersymmetric black hole solutions  in terms of a  `thermodynamic energy' as in \cite{Cvetic:2005zi} and then take a supersymmetric limit, and then choose the local parameters in these metrics so that a soliton geometry, rather than a black hole, is obtained.  However, the Ashtekar-Magnon definition appears more natural here, particularly since solitons have no horizon and hence no associated temperature \footnote{Nonetheless, solitons are still thermodynamically relevant.  One can derive a `soliton mechanics', i.e. a Smarr-type relation and variation formulae for both globally and locally Anti-de Sitter solitons~\cite{Mbarek:2016mep, Andrews:2019hvq, Durgut:2022xzw}. Moreover,  solitons arise in Hawking-Page type phase transitions between black holes and solitons with the same conformal boundary~\cite{Surya:2001vj, Durgut:2022xzw}.}.  As further evidence in support of identifying $\mathbf{M}_{\text{AD}}$ with the mass of the soliton spacetimes, consider the holographic stress tensor approach \cite{Balasubramanian:1999re, deHaro:2000vlm} . Adapted to the present setting, one considers a dual CFT on $\mathbb{R} \times L(p,1)$ in a spacetime with conformal metric \eqref{boundary} and computes the expectation value of the stress tensor given by (in units where the gravitational constant $G=1$)
\begin{equation}
\langle T_{\mu \nu} \rangle =\lim_{R\to \infty} \frac{R^2}{8 \pi \ell^2} \left[-(K_{\mu\nu} - \text{Tr}_{\bar{h}} K \bar{h}_{\mu\nu}) - \frac{3}{\ell} \bar{h}_{\mu\nu} + \frac{\ell}{2} \bar{G}_{\mu\nu} \right]
\end{equation} where $(\bar{h},K)$ are the (Lorentzian) metric and extrinsic curvature associated to the surfaces $R =$constant in the asymptotically static chart \eqref{AGAdS}, and $\bar{G}_{\mu\nu}$ is the Einstein tensor of $\bar{h}$.  In particular, we choose the outward pointing normal so that $K_{\mu \nu}  = \dot {\bar{h}}_{\mu \nu} / (2 g_{RR})$ where the overdot $\dot{}$ denotes a derivative with respect to $R$.  The holorgraphic energy is then obtained by integrating the function $T_{TT}$ over the conformal boundary with metric \eqref{boundary}. One obtains
\begin{equation}
\mathbf{E} = \int_{L(p,1)} T_{TT} \; \text{dvol}(\bar{h}) = \frac{\ell^2 \pi (-79 + 96p + 24p^2 - 16 p^3)}{864 p}
\end{equation} which is strictly negative for $p \geq 3$. This energy can be expressed in terms of the Ashtekar-Magnon mass as
\begin{equation}
\mathbf{E} = \mathbf{M}_{\text{AD}} + \frac{3 \ell^2 \pi}{32p}. 
\end{equation} The second term is recognized to be the Casimir energy of the CFT on $\mathbb{R} \times L(p,1)$; this is the energy of the AdS$_5 / \mathbb{Z}_p$ orbifold (it has an orbifold singularity at the fixed point of the $SO(4)$ action). Note that $\mathbf{M}_{\text{AD}} < 0$, so the presence of the (globally smooth) soliton \emph{lowers} the energy with respect to the orbifold vacuum geometry. 

Finally, we note that the non-contractible $S^2$ at $r = r_0$ carries a `dipole charge' $\mathcal{D}$ which is physically interpreted as a flux of $F$ that prevents its collapse. In particular note that $F$ is not globally exact, and the dipole charge is 
\begin{equation}
\mathcal{D} = \frac{1}{4\pi} \int_{S^2} F = \frac{\ell (p-2)}{4\sqrt{3}}.
\end{equation} The charge is not conserved in the sense that it is only non-zero if the 2 surface on which it is defined encloses the bubble. 

\subsubsection{Solutions with non-equal charges} A second set of smooth solutions can be obtained by choosing the dimensionless integration constant vector to take the form $y_0^i = (y_0, y_0, \beta y_0)$ for $\beta \in \mathbb{R}$. We then have for the charge parameters
\begin{equation}
\mathfrak{q}_1 = \mathfrak{q}_2 = -\frac{4 \beta y_0^2}{g^4 a_0 x}, \qquad \mathfrak{q}_3 = -\frac{4 y_0^2}{g^4 a_0 x}
\end{equation} where, as above, $x = r_0^2 g^2$ is dimensionless.  The constraints \eqref{ycons} immediately can be used to find 
\begin{equation}
y_0 = \frac{g^2 (a_1 r_0^2 - a_0)}{2(2+\beta)}
\end{equation} and using the regularity condition \eqref{regularity} to eliminate $a_1$ as well as \eqref{consSTU} one can follow the procedure described above to eliminate $a_0$ and obtain a polynomial equation for $x$ with coefficients depending on the parameter $\beta$ and the natural number $p$:
\begin{equation}\label{pol}
\begin{aligned}
&4(\beta-1)^3 x^3 - \left((\beta-1)^2 p(p+2) - 7\beta^2 + 22 \beta +4)\beta + 8\right) x^2 \\
&+ (p-2)(4p + \beta(\beta^2 (p-1)^2 -4 + 2p (p+2) - 2\beta(1 +p(p-4))) x - p \beta(p-2)^3 =0.
\end{aligned}
\end{equation} For $\beta=1$ this reduces to the quadratic that determines $x$ in the minimal theory \eqref{minimal}. We also note that the case $\beta =0$ leads to a solution for which $p \geq 3$ and $\mathbf{Q}_1 = \mathbf{Q}_2 \neq 0, \mathbf{Q}_3 =0$. Although this special case appears pathology-free, we find that it has $\mathbf{M}_{\text{AD}} =0$ so we will not pursue it further.

Rather than attempting an exhaustive analysis of the full space of solutions of \eqref{pol}, we will confine ourselves to some simple illustrative example. 
An obvious question is whether one can obtain asymptotically globally AdS$_5$ solitons ($p=1$) which, as we explained above, cannot exist for $\beta =1$.  We have not yet been able to find any smooth examples with simple values for $\beta$ (one finds the function $f$ has zeroes for $r > r_0$). However, a systematic investigation may lead to examples. 
 Consider the case $\beta = 2, p =3$.  Then the cubic \eqref{pol} factors as $2(2x - 3)(1 - 18x + x^2)$ producing solutions $x_1 = 3/2, x_2 = 9 + 4\sqrt{5}, x_3 = 1/x_2$.  Consider $x= x_1$. Then we have $a_0 = 3/(8g^2)$, $a_1 = 5/4$, and $y_0 = 3/16$. This leads to 
\begin{equation}
\mathfrak{q}_1 = \mathfrak{q}_2 = -\frac{1}{2g^2}, \qquad \mathfrak{q}_3 = -\frac{1}{4g^2}.
\end{equation} Then we may write
\begin{equation} 
f^3 = \frac{r^6}{\left( r^2 - r_0^2 + 2\ell^2)^2(r^2 - r_0^2 + \frac{5 \ell^2}{4} \right)} 
\end{equation} which is obviously positive for $r  > r_0 = \sqrt{3/2} \ell$.  The solution has a mass $\mathbf{M}_{\text{AD}} = -\pi \ell^2 / 48$ and $\mathbf{Q}_1 = \mathbf{Q}_2 = -3 \mathbf{Q_3}/2 = 3\pi \ell^2/32$.  Repeating this procedure for the case $\beta =4, p =3$ leads to a well behaved solution of \eqref{pol} with $x_\pm = 2/3 \pm \sqrt{11/27} > 0$. The resulting solutions for either root leads to regular solitons with non vanishing mass $\mathbf{M}_{\text{AD}}$, angular momentum $\mathbf{J}$ and charges $\mathbf{Q}_i$. 

\section{Discussion}
We have constructed asymptotically AdS$_5 / \mathbb{Z}_p$ supersymmetric gravitational soliton solutions of five-dimensional gauged supergravity coupled to an arbitrary number of vector multiplets.  The solutions are globally stationary and admit an $SU(2) \times U(1)$ isometry group. The local solutions are characterized by $N+1$ continuous parameters $(r_0, U_0^I)$. The local solutions extend to globally defined metrics provided these parameters satisfy an additional two constraints for a given $p \in \mathbb{N}$ which determines the topology of the conformal boundary.  We have investigated in detail a subset of these local solutions for which we can explicitly solve these constraints for $p \geq 3$ and showed there are examples beyond the minimal theory.  Given their similarity to the Eguchi-Hanson-AdS$_5 / \mathbb{Z}_p$ vacuum solitons, we could identify these solutions as supersymmetric generalizations that carry, in addition to negative mass relative to the AdS orbifold background, non-vanishing angular momentum and charge. 

We close with a brief discussion of some further problems which arise as a consequence of our work.  A natural question to investigate is the question of stability. Famously, robust numerical investigation have provided strong evidence that global AdS is nonlinearly unstable~\cite{Bizon:2011gg, Dias:2011ss} to the formation of black holes as energy tends to get concentrated to shorter scales.  One might expect supersymmetric solitons in AdS to suffer from a similar instability.  Dold has rigorously established that the maximal development of  $SU(2) \times U(1)$-invariant initial data sufficiently close to Eguchi-Hanson-AdS$_5 / \mathbb{Z}_p$ do not form future horizons. This suggests that the endpoint of the evolution (assuming it settles to a static spacetime with the same conformal boundary) would be a spacetime containing a naked singularity.

 More recently, mode solutions of the linear scalar wave equation on the Eguchi-Hanson-AdS$_5/ \mathbb{Z}_p$ soliton were analyzed~\cite{Durgut:2022xzw}. It was shown that, similar to AdS, the geometry admits a normal mode spectrum, so that scalar modes neither grow nor decay in time. These two stability results are consistent, as \cite{Durgut:2022xzw} is purely a linear result.  It would be interesting to extend both of these analyses to the stationary, supersymmetric solitons constructed here to determine whether there are obstructions to horizon formation and/or a normal mode spectrum.  A related challenging problem would be to study linearized gravitational perturbations of the background. Since the solutions have $SU(2) \times U(1)$ symmetry, one could decompose metric and Maxwell field perturbations using the strategy developed in~\cite{Murata:2008xr}.
 
 Our preliminary analysis at least reveals that there are no ergoregions (with respect to either the stationary Killing fields $\partial_t$ or $\partial_T$), which are known to be another channel for instabilities. As mentioned in the Introduction, this is in stark contrast to the 1/2-BPS supersymmetric solitons constructed in~\cite{Durgut:2021rma} which actually contain evanescent horizons (the supersymmetric Killing vector field becomes null on a co-dimension two timelike surface) which provide a geometric mechanism for instability.  The absence of such tapping mechanisms suggests that the supersymmetric solitons constructed here could be nonlinearly stable.

\paragraph{Acknowledgements} We are grateful to James Lucietti for useful comments. HKK acknowledges the support of the NSERC Grant RGPIN-2018-04887.

\end{document}